\documentclass[]{raa}            
\usepackage{graphicx,times}
\usepackage{natbib}

\providecommand{\bm}[1]{\mbox{\boldmath$#1$\unboldmath}}

\begin{document}

   \title{Research of the active reflector antenna using laser angle metrology
system
\,$^*$
\footnotetext{$*$ supported by the National Natural Science
Foundation of China and the Knowledge Innovation Program of Chinese Academy
of Sciences.}
}

   \volnopage{Vol.0 (200x) No.0, 000--000}      
   \setcounter{page}{1}          

   \author{Yong Zhang
      \inst{1,2}
   \and Jie Zhang
      \inst{1,2,3}
   \and Dehua Yang
      \inst{1,2}
   \and Guoha Zhou
      \inst{1,2}
   \and Aihua Li
      \inst{1,2}
   \and Guoping Li
      \inst{1,2}
   }

   \institute{National Astronomical Observatories/Nanjing Institute of Astronomical
Optics {\&} Technology, Chinese Academy of Sciences, Nanjing 210042, China; {\it jzhang@niaot.ac.cn}\\
        \and
             Key Laboratory of Astronomical Optics {\&} Technology, Nanjing Institute
of Astronomical Optics {\&} Technology, Chinese Academy of Sciences, Nanjing
210042, China\\
        \and
             Graduate University of Chinese Academy of Sciences, Beijing 100049, China\\
   }

   \date{Received~~2009 month day; accepted~~2009~~month day}

\abstract{ Active reflector is one of the key technologies for
constructing large telescopes, especially for the millimeter/sub-millimeter
radio telescopes. This article introduces a new efficient laser angle
metrology system for the active reflector antenna of the large radio
telescopes, with a plenty of active reflector experiments mainly about the
detecting precisions and the maintaining of the surface shape in real time,
on the 65-meter radio telescope prototype constructed by Nanjing Institute
of Astronomical Optics and Technology (NIAOT)\footnote{
http://65m.shao.cas.cn/.}. The test results indicate that the accuracy of
the surface shape segmenting and maintaining is up to micron dimension, and
the time-response can be of the order of minutes. Therefore, it is proved to
be workable for the sub-millimeter radio telescopes.
}

   \authorrunning{Yong Zhang, Jie Zhang, Dehua Yang, Guohua Zhou, Aihua Li, Guoping Li }            
   \titlerunning{Research of the active reflector antenna using laser angle metrology
system }  

   \maketitle

\section{Introduction}           
\label{sect:intro}

Almost all the inspiring achievements in radio astronomy are due to the
observing equipments and technologies, which work on new band, provide newer
detecting precision or higher resolution of time and space (Xiang 1990). In
recent decades, many centimeter/millimeter/sub-millimeter radio telescopes
have been built, such as radio telescope Effelsberg in German\footnote{
http://www.mpifr-bonn.mpg.de/english/radiotelescope/index.html.}, Green Bank
Telescope (GBT) in the U.S.\footnote{
https://science.nrao.edu/facilities/gbt}$^{ }$and so on. These telescopes
have a extremely large aperture composed by numerous panels, GBT is composed
by 2004 panels as a example (Mckinnon 2010, in Synthesis Summer School). On
the process of the installation and the formal operation, gravitational,
thermal and wind-driven effects would induce the deformation of a radio
antenna's main reflector. In order to permit the high performance
observation at any time, besides the design of improved and enough stiff
antenna mechanical support, active reflector antenna technology is now
developed as a key (Zhang et al. 2010a).

So far the measurements of the active reflector mainly have the following
three types: The first is the rangefinder based technique (Levy 1996), like
the laser metrology system adopted by the GBT as a successful design (Parker
1997). The data analysis is pretty tedious since that the final accuracy
depends on the sophistication and the data volume, thus the time-response of
this technique is of the order of hours for achieving the limited accuracy
of 100 microns. The second technology is known as the microwave holography
(Kesteven 1994; Wang {\&} Yu 2007). It uses the Fourier transform pair
relation between antenna aperture field and the far field pattern, which can
only be found on some specified elevation angles. The accuracy is quite good
(of the order of tens of microns), however the time-response depending on
the desired surface resolution can be of the order of many hours. The third
one is the photogrammetric system (Fraser 1986; Wang et al. 2007). It
reaches the accuracy of tens of microns with a laborious set-up, and
time-expensive data acquisition and analysis (usually many hours).

All of the above measurements share the same shortcoming that the precisions
cannot satisfy the sub-millimeter radio telescope's active reflector
antenna, and the deformation correction cannot be accomplished in real time.
Here we introduce a laser-angle-metrology-based active reflector antenna,
which approaches the accuracy of microns and has a useful time-response for
the active optics close loop (minutes as an order of magnitude). Through the
experiments it has been successfully proved.

\section{Detection principle}
\label{sect:principle}

In optical telescope, the Shack-Hartmann wave-front sensor is widely used.
The laser angle metrology system can be seen as a simplified
Shack-Hartmann-type wave-front sensor. It images the laser spot from a paper
screen onto the video CCD camera, which are defined by the laser transmitter
modules installed and well aligned on each panel. By calculating the spot
position it can deduce the normal angle deflection and axial displacement of
the panel (Zhang et al. 2010a,b).

Laser angle metrology system includes angle measurement and range
measurement. A sketch of the angle measuring is drawn on figure 1. The panel
tilt ($\Delta \theta )$ leads to the position change of the laser spots
(D). Using the formula $\Delta \theta= \arcsin{\frac{D}{L}}$, one can easily obtain the tilt of the panel,
thereby can achieve the correction of the normal-direction deflection. Range
measuring is sketched in the following figure 2. Two different laser spots
on the screen are formed by two laser beams of a certain distance, if the
panel moves in the axial direction, the distance between the two laser spots
would change as well. The axial displacement of the panel is given in the
formula $\frac{\Delta s}{S}= \frac{l}{L}$, where S is the distance between the two laser transmitter modules.

\begin{figure}[htbp]
   \centering
   \includegraphics[scale=0.5]{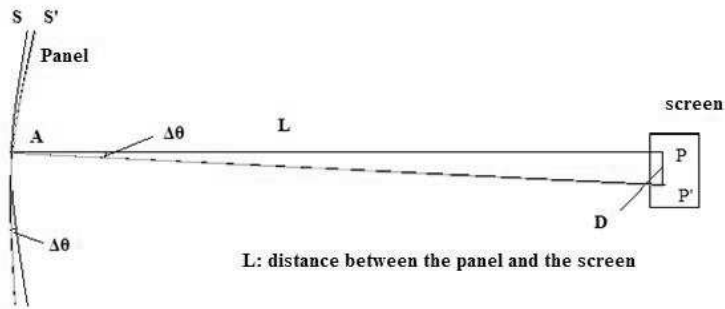}
   \caption{The illustration of the angle measurement.}
   \label{Fig1}
   \end{figure}

\begin{figure}[htbp]
   \centering
   \includegraphics[scale=0.5]{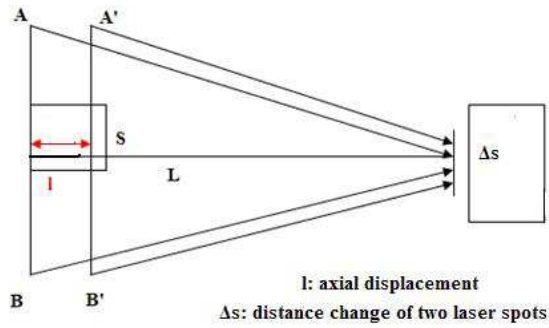}
   \caption{The illustration of the ranging based on angle measurement.}
   \label{Fig2}
   \end{figure}

Because the test of the normal angle of each panel is influenced by the
laser transmitter module's initial installation condition, especially
pointing precision, the laser angle metrology system is not suitable for the
initial calibration of the surface shape. But, it is simple and of great use
for maintaining the active surface shape. The whole system is easily to
construct and automatically to implement, with high efficiency and low cost
(Zhang et al. 2010a).

\section{System implementation}
\label{sect:sys}

The system is implemented based on the 65-meter radio telescope prototype
constructed by NIAOT (figure 3). It has four panels segmented on the five
support points like Point A. Under the support points are mechanical
displacement actuators, which the panel shares with the neighboring ones at
the corner. Each panel has two distanced lasers on the position like 1 and
2. A CCD camera system as a detector is placed in the front of the panels to
guarantee the accuracy considering the limited short distance between the
panels and the detector position. The actual experiment system is shown in
figure 4.

\begin{figure}[htbp]
   \centering
   \includegraphics[scale=0.5]{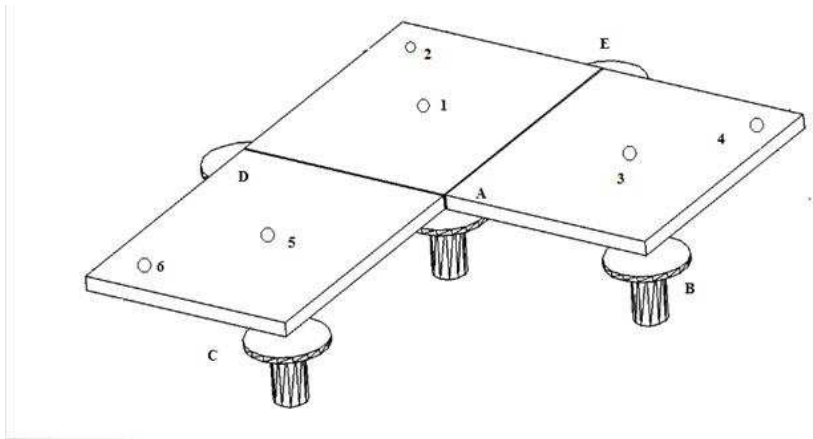}
   \caption{65-meter radio telescope prototype system with four panels.}
   \label{Fig3}
   \end{figure}

\begin{figure}[htbp]
   \centering
   \includegraphics[scale=0.25]{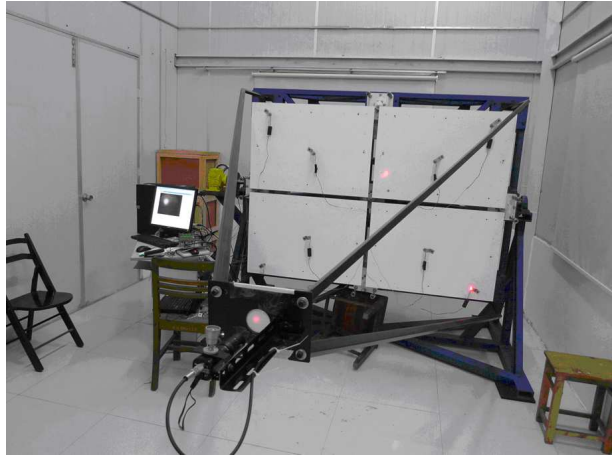}
   \caption{The actual equipment of the laser angle metrology system.}
   \label{Fig4}
   \end{figure}

\newpage

\section{Test results}
\label{sect:test}
To ensure a radio telescope working with diffraction-limited performance,
the accuracy (root mean square) of the antenna surface deformation should be
less than 1/40 wavelength. For a sub-millimeter antenna (0.2mm), the
accuracy of the surface shape is about 5 microns (RMS), and 10 microns (RMS)
are usually the specifications of all constructed sub-mm radio telescope.

Whether the laser angle metrology system is available for correcting the
deformation of surface shape of a sub-millimeter radio telescope or not, depends on
the detecting precisions satisfied with the required 5 microns. Therefore,
we measure the precision of the laser spot detection at length, then test
the accuracies of the surface shape segmenting and maintaining.

\subsection{The precision of the laser spot detection}
\begin{figure}[htbp]
   \centering
   \includegraphics[scale=0.35]{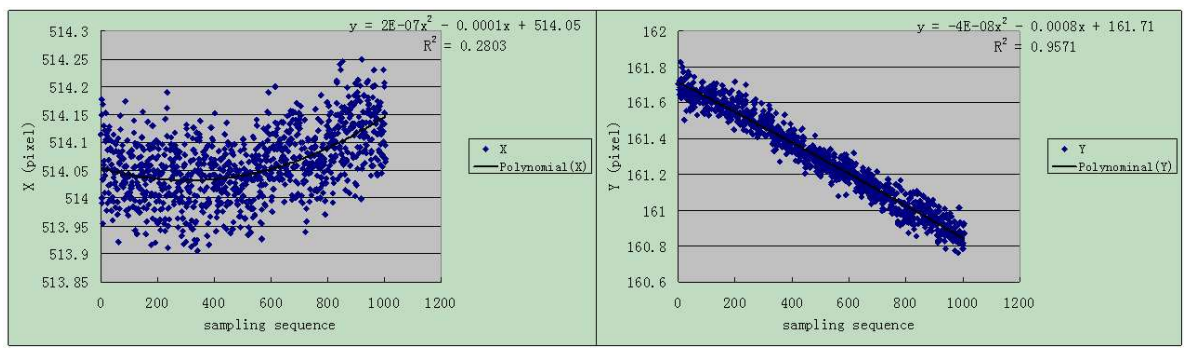}
   \caption{The X, Y value of the sampling laser spot.
(27${^\circ}$, the brightness of the system display is 100 and the contrast
of display is 150)
}
   \label{Fig5}
   \end{figure}
As the sampling laser spot shown in figure 5, the positions of the detecting
spot change over time with a second order trend. This trend is possibly
caused by the heating and gravity deformation of the laser transmitter
module. The precision is strongly influenced by it, especially the
peak-to-valley (PTV) value, see figure 6(1), the PTV value is down to 0.6
pixel.

In the following experiments of angle and range measurement, considering the
trend, the data we adopt are the average of a hundred sampling points. So
the real precision of the laser spot detection is of the residual that
obtained from the sampling 100 spots removing the second order trend. As
shown in figure 6(2), the precision can be up to 0.02 pixel RMS and 0.06
pixel PTV.

\begin{figure}[htbp]
   \centering
   \includegraphics[scale=0.35]{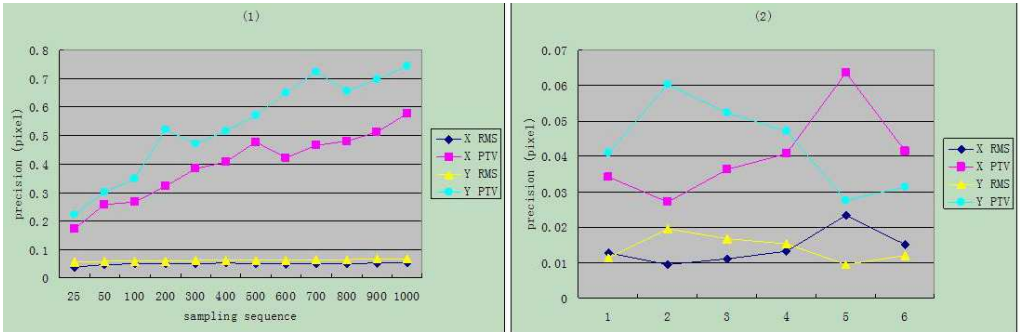}
   \caption{(1) The precision of the laser spot detection; (2) The precision after removing the trend.)
}
   \label{Fig6}
   \end{figure}

Then we discuss the accuracy of the position of the laser spot. As the
schematic diagram of the laser spot in figure 7, Point P is the average of
the sampling points, P' is any point of the detection, $\Delta P=\vert
P'-P\vert $ represents the distance between them. Based on Gaussian model,
the discrete diameter of the laser spot (80{\%} encircled energy of the
laser spot distribution) characterizes the accuracy of the position. The
accuracy from experiments is about 0.2 pixel, as shown in figure 8.

\begin{figure}[htbp]
   \centering
   \includegraphics[scale=0.4]{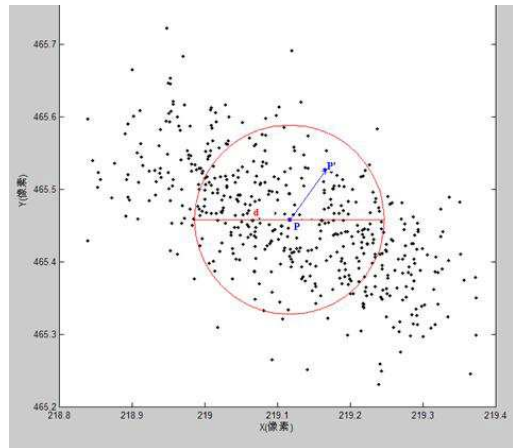}
   \caption{The schematic diagram of the laser spot (unit: pixel).
(25${^\circ}$, the brightness of the system display is 150 and the contrast
of display is 150)
}
   \label{Fig7}
   \end{figure}

\begin{figure}[htbp]
   \centering
   \includegraphics[scale=0.4]{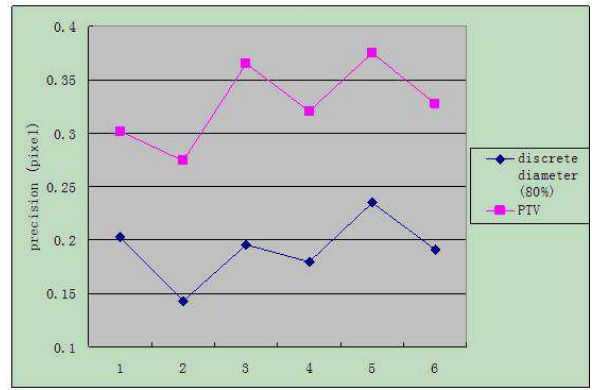}
   \caption{The accuracy of the laser spot position.}
   \label{Fig8}
   \end{figure}

\newpage
\subsection{The precisions of angle measuring and surface shape segmenting}

The precision of the angle measuring can be directly derived from the
precision of the laser spot detection. In our experiment, 764 pixels of CCD
camera can receive a laser spot of the size of 50mm, the distance between
the panel and the screen is 2315mm. Calculating from the formula $\Delta \theta= \arcsin{\frac{D}{L}}$ discussed in section 2, the precision of the angle measuring is up to 0.11 arc sec.

Integrate all the laser spots from the lasers on four panels to the same
coordinate system, with zero time for the origin, thus the position accuracy
of the laser spot integrated indicates the precision of the surface shape
segmenting. See figure 9, the diameter of the circle is the discrete
diameter of the laser spot (80{\%} encircled energy of image spot
distribution) based on Gaussian model. Fit the surface shape segmenting
ANSYS-based with the normal tilt of each panel at certain moments, as shown
in figure 10. It is obviously that the greater the laser patches dispersed,
the worse the deformation of the surface shape. The precision of the surface
shape segmenting is up to 2$\mu $m (RMS), which can easily meet the
constraint of the sub-millimeter antenna.

\begin{figure}[htbp]
   \centering
   \includegraphics[scale=0.4]{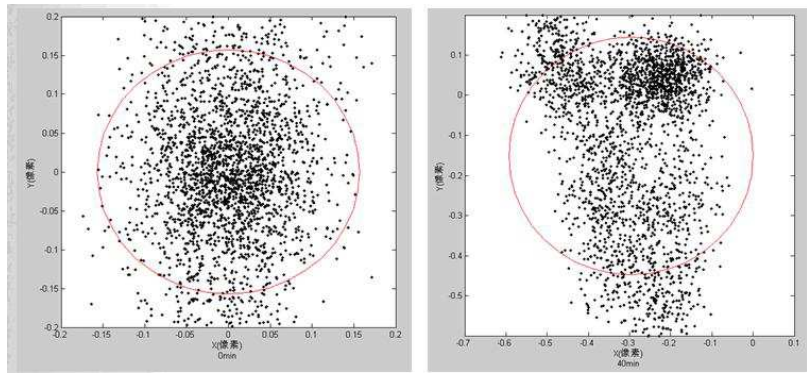}
   \caption{The schematic diagram of the laser spot integrated (unit: pixel).}
   \label{Fig9}
   \end{figure}

\begin{figure}[htbp]
   \centering
   \includegraphics[scale=0.3]{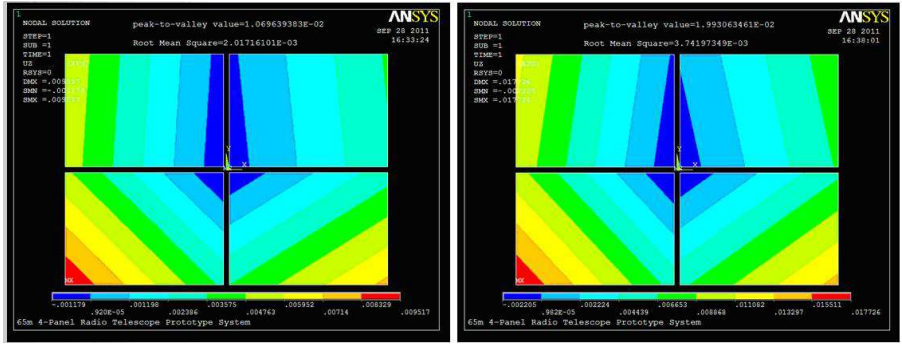}
   \caption{The ANSYS-based surface shape segmention related to figure 9 (unit: mm).}
   \label{Fig10}
   \end{figure}

\newpage
\subsection{The maintaining of the active reflector surface shape}

We correct the deformation of the surface shape using the stiffness matrix
of this system, which is a 16X5 matrix according to the equipment and
obtained by testing repeatedly. For a surface having random disturbance, the
steps below have been taken: First, we figure out the normal angle
deflections of the panels, then the displacement calibrations of the
actuators by the stiffness matrix. After the movement of the actuators, we
test the panel tilts as a iteration. The time-response of one cycle is less
than 5 minutes, including 2 minutes of data acquisition and about 2 minutes
of the separately handy calculation of each panel's deflection. With a
complete automatic experiment system, the time-response may be within 3
minutes.

The correcting test is shown in figure 11. After three steps, the
deflections of the panels tend to be stable at the range of 5 arc sec, which
is equal to the detecting accuracy of the stiffness matrix as a result of
systematic residuals. As shown in figure 12, the ANSYS-based surface shape
segmenting indicates that, with three steps the accuracy of the surface
shape segmenting is from 1mm down to 4$\mu $m (RMS). The time-cost of the
correcting procedure is amazingly short (shorter than 15 min), and the precision totally
satisfies the demand of a sub-millimeter antenna. The feasibility of the
laser angle metrology system for the sub-millimeter radio telescope is
successfully proved.

\begin{figure}[htbp]
   \centering
   \includegraphics[scale=0.4]{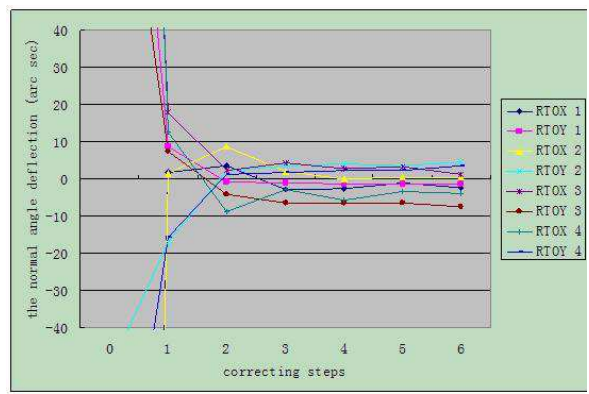}
   \caption{The normal angle deflection of each panel.}
   \label{Fig11}
   \end{figure}

\begin{figure}[htbp]
   \centering
   \includegraphics[scale=0.3]{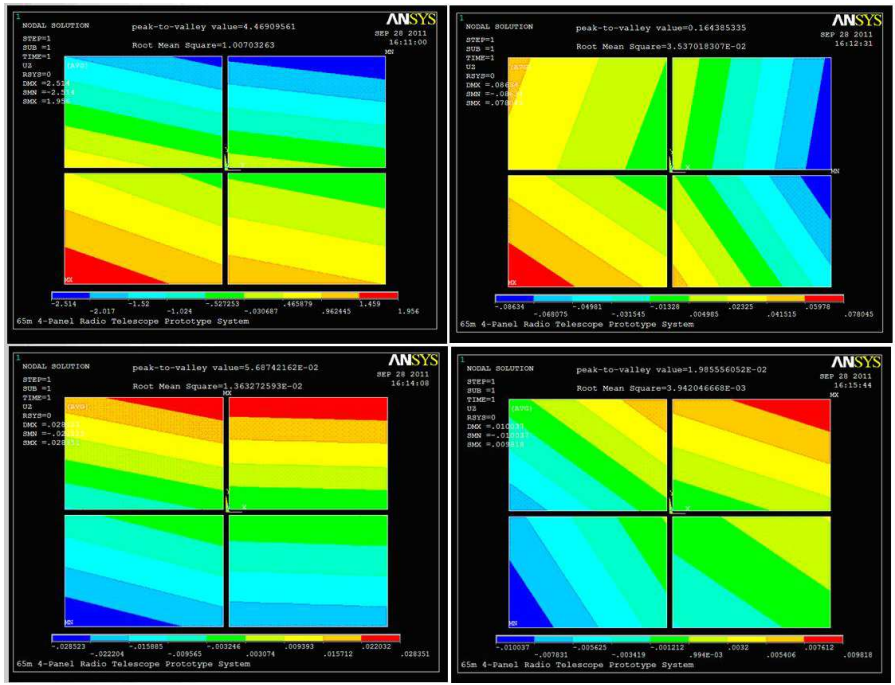}
   \caption{The ANSYS-based surface shape segmenting using 16X5 stiffness
matrix (unit: mm).(The left upper: shape segmenting after a random disturbance; the right
upper: after one correcting step; the left lower: after two steps; the right
lower: after three steps)}
   \label{Fig12}
   \end{figure}

\newpage
\subsection{The precision of range measuring}

According to the formula $\frac{\Delta s}{S}= \frac{l}{L}$ in section 2, the precision of the range measuring
is represented by the precision of $\Delta $s. Measure the distance changes
of the image patches positions of four panels respectively. The accuracy,
shown in figure 13, reaches 10$\mu $m dimension (RMS). The reason why the
accuracy is limited is the gravity deformation of the laser transmitter
module. As shown in figure 14, the Y values of the laser patches decrease as
time, that is to say, the positions of the image patches shift down as time
which apparently influences the detecting of $\Delta $s badly. The
experiment with improved devices shows that the accuracy can be up to 5$\mu
$m (RMS) in figure 15.

\begin{figure}[htpb]
   \begin{minipage}[t]{0.495\linewidth}
   \centering
   \includegraphics[scale=0.35]{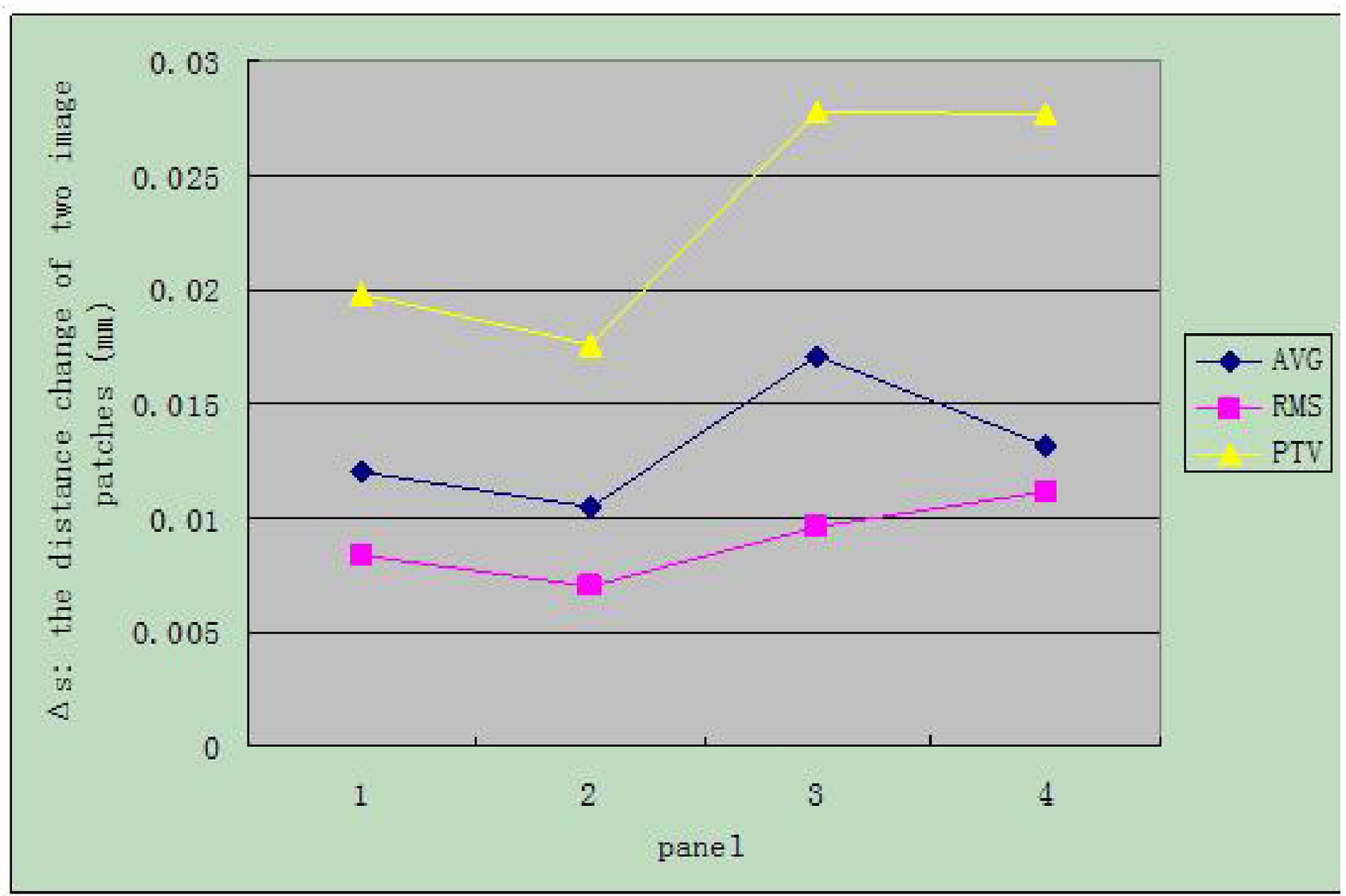}
   \caption{The precision of $\Delta $s for four panels.}
   \end{minipage}%
  \begin{minipage}[t]{0.495\textwidth}
  \centering
   \includegraphics[scale=0.35]{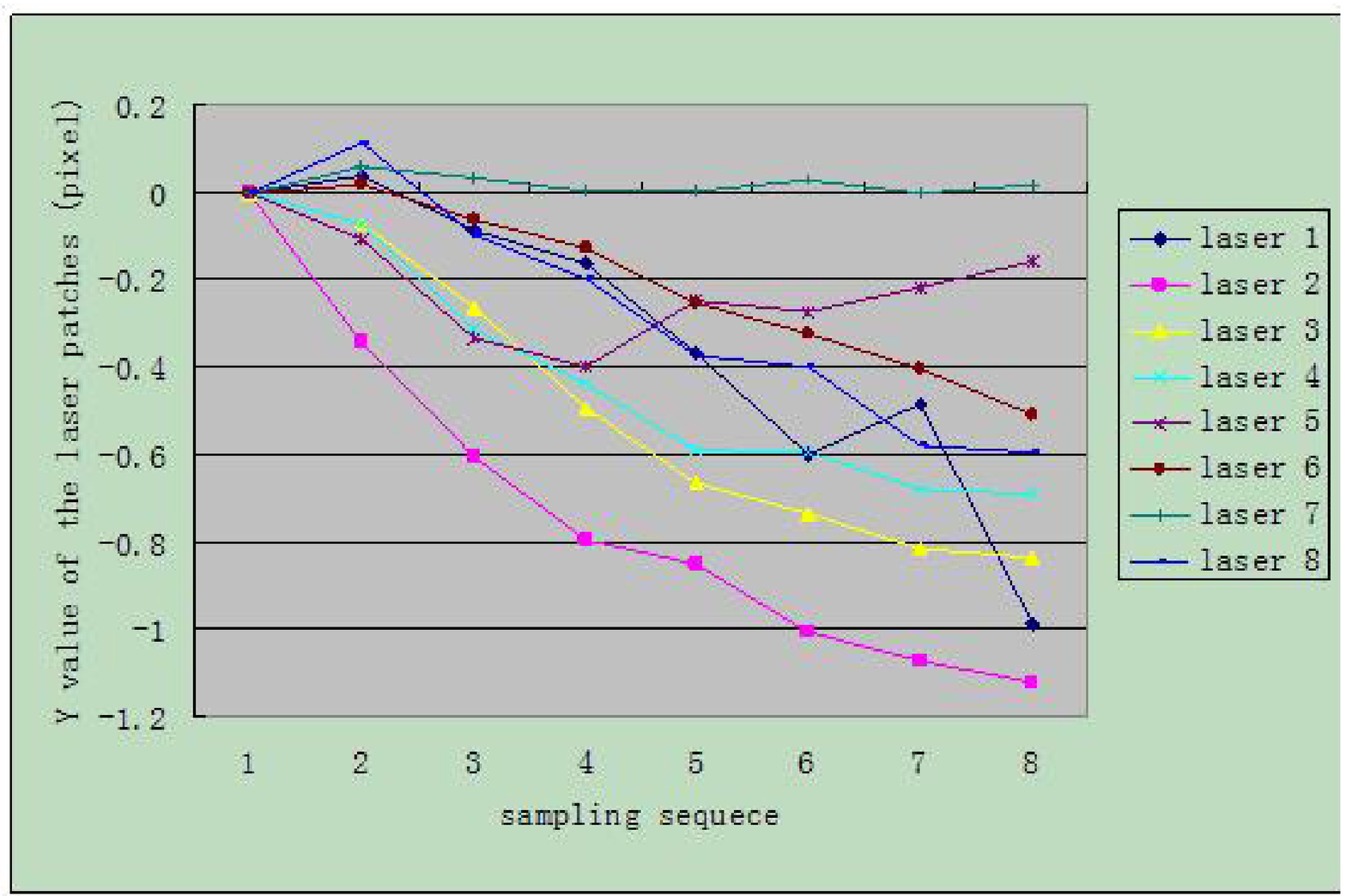}
   \caption{The Y values of the laser patches decrease as time.}
   \end{minipage}%
  \label{Fig13}
   \end{figure}

\begin{figure}[htbp]
   \centering
   \includegraphics[scale=0.4]{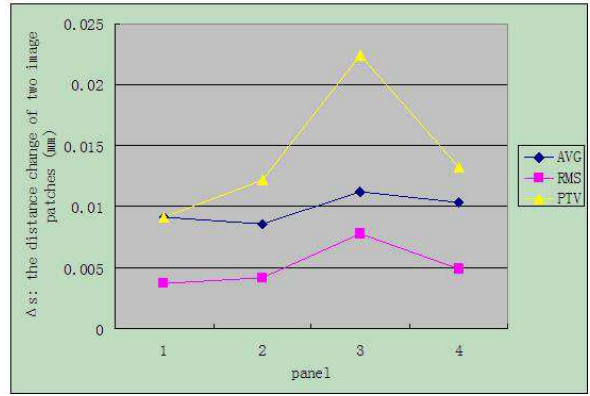}
   \caption{The precision of $\Delta $s for four panels with improved devices.}
   \label{Fig14}
   \end{figure}

\newpage
\subsection{The elimination of the influence of lateral deformation of the reflector}

Firstly the lateral deformation of a single panel is very small, accords with the technical requirements of the panel installation guaranteed by finite element analysis and actual installation and alignment. In the next place, the panel, as a laser measuring basis and a locator of the optic axis of the system, is sensitive to the lateral displacement (in proportion of 1:1), after correcting the normal angle deflection and axial displacement. Thus, the precision of lateral displacement of the reflector can be straightly given by the precision of the laser spot detection (0.02 pixel RMS in section 4.1). In our experiment, the detection precision of lateral displacement of the reflector is  50mm*0.02pixel/764pixel=1.3$\mu $m (the data are given in section 4.2).

The influence of the lateral deformation would be eliminated by the secondary reflector and feed system of the radio telescope, which are normally high-accuracy adjustable.

\section{Conclusions}
\label{sect:conclusion}

With the continuously extension from millimeter to sub-millimeter wavelength
on radio astronomy, the demand of the radio telescope becomes more and more
strict. Through a plenty of experiments on 65-meter radio telescope
prototype, it is obtained that the precision of the laser spot detection is
0.02 pixel, the precision of the surface shape segmenting is up to 2$\mu $m,
the precision of range measuring is up to 5$\mu $m, and the accuracy of the
surface shape maintaining achieves 5$\mu $m (RMS) with a time-response less
than a quarter. All the results above strongly prove that the laser angle
metrology system is available for the sub-millimeter radio telescope
antenna, both in accuracy and time-response. This technology can be applied
to the reconstruction of the active reflector antenna in China, and it will
play a promoting role in developing the new area of sub-millimeter radio
astronomy.

\begin{acknowledgements}
This work is supported by the National Natural
Science Foundation of China (Granted NO. 10703008, 11073035 and 10833004)
and the Knowledge Innovation Program of Chinese Academy of Sciences (Granted
NO. KJCX2-YW-T17).
\end{acknowledgements}

\label{lastpage}

\end{document}